\documentclass{article}
\usepackage[utf8]{inputenc}
\usepackage[a4paper, total={5.5in, 9.7in}]{geometry}

\usepackage{authblk} % for author affiliations
\usepackage{graphicx} % for inserting figures
\usepackage{physics}
\usepackage{amsmath}
\usepackage{mathtools}
\usepackage{verbatim} % package for using the "comment" syntax
\usepackage{float} % for using [H] to place the figure exactly at this position
\usepackage{xcolor} % for coloring text
\usepackage{cite} % for citing to be shown as a range like [1-5]
\usepackage[resetlabels,labeled]{multibib}

\newcommand{\supplementaryinformation}{%
  \setcounter{section}{0}% Reset section counter
  \let\oldthesection\thesection% Capture section numbering scheme
  \renewcommand{\thesection}{S\oldthesection}% Prefix section number with S
  
  \setcounter{figure}{0}% Reset figure counter
  \let\oldthefigure\thefigure% Capture figure numbering scheme
  \renewcommand{\thefigure}{S\oldthefigure}% Prefix figure number with S
  
  \setcounter{equation}{0}% Reset equation counter
  \let\oldtheequation\theequation% Capture equation numbering scheme
  \renewcommand{\theequation}{S\oldtheequation}% Prefix equation number with S
  
  \part*{Supplementary Information}% Set supplementary information
} %

\title{Mapping single-shot angle-resolved spectroscopic micro-ellipsometry with sub-5 microns lateral resolution}
\author[ \hspace{-0.8ex}]{Ralfy Kenaz}
\author[*]{Ronen Rapaport}
\affil[ \hspace{-1ex}]{\small{Racah Institute of Physics, The Hebrew University of Jerusalem, Jerusalem 9190401, Israel}}
\affil[*]{\small{ronenr@phys.huji.ac.il}}

\date{}

\begin{document}

\maketitle

\begin{abstract}

Spectroscopic ellipsometry is a widely used optical technique both in industry and research for determining the optical properties and thickness of thin films. The effective use of spectroscopic ellipsometry on micro-structures is inhibited by technical limitations on lateral resolution and data acquisition rate. Here we introduce a spectroscopic micro-ellipsometer (SME), capable of measuring spectrally resolved ellipsometric data at many angles of incidence in a single-shot with a lateral resolution down to 2 microns. The SME can be easily integrated into generic optical microscopes by addition of a few stock optics. We demonstrate complex refractive index and thickness measurements by the SME which are in excellent agreement with a commercial spectroscopic ellipsometer. As an application for its accuracy and high lateral resolution, the SME can characterize the optical properties and number of layers of exfoliated transition-metal dichalcogenides and graphene, for structures that are a few microns in size.

\end{abstract}

\section{Introduction}

Ellipsometry is a powerful optical technique for thin film characterization, based on measuring the change in light polarization upon oblique reflection. Its non-destructive nature, high accuracy, simplicity and availability make ellipsometry an essential tool in various fields of industry and research, such as semiconductors \cite{Zollner2013SpectroscopicIndustry,Orji2018MetrologyDevices,Kwon2022Microsphere-assistedDevices}, photovoltaics \cite{2018SpectroscopicCharacterization,Yan2020DeterminationFilms}, materials characterization \cite{Wahaia2017EllipsometryCharacterization,Martin2019ReliableInvited}, optical coatings \cite{Hilfiker2019SpectroscopicCoatings,Abbasi-Firouzjah2020DepositionPolymerization}, two-dimensional materials \cite{Yoo2022SpectroscopicHeterostructures, Ermolaev2020BroadbandMoS2}, flat panel displays \cite{Woollam1994SpectroscopicMaterials, Gaillet2007OpticalEllipsometry, Bohorquez2022SpectroscopicDeposition}, organic films and surfaces \cite{2018EllipsometryFilms, Pinto2022SpectroscopicHybridization}, antifouling coatings \cite{Gnanasampanthan2021EffectCoatings, Yu2021Layer-by-LayerProperties}, biological materials \cite{Arwin2011ApplicationMaterials, Kariper2021ADetection} and many more.

In ellipsometry, measured data is fitted to a relevant model for extracting optical properties or thickness information. Increasing the measurement data points by spectroscopic and angle-resolved acquisition is crucial for minimizing the local-minima problem \cite{Woollam1999OverviewApplications, LingjieLi2021SpectroscopicEllipsometry, Choi2020Single-shotEllipsometry} and increasing the sensitivity and accuracy of the fit parameters. Spectroscopic ellipsometry (SE) uses broadband illumination and obtains spectrally resolved ellipsometric data at a single angle of incidence (AOI) in a single-shot. The AOI can be varied between measurements by manual or automatic alignment of the mechanical arms of the ellipsometer instrument. Besides providing spectral information, SE improves the accuracy and data acquisition rate when compared to single-wavelength ellipsometry. This significant advantage makes SE a standard method among the polarization-dependent optical techniques for investigation of optical properties \cite{Schubert2005TheoryEllipsometry}.

On the flip side, SE suffers from low lateral resolution due to its off-axis configuration. Conventional SE uses quasi-collimated beams resulting in elliptical and millimeter-scale beam shapes on the sample. In order to improve the lateral resolution, focused-beam SE adds low numerical aperture (NA) objective lenses before and after the sample, reducing the spot size down to tens-of-microns \cite{Barton2006EllipsometerBeams,SangJunKim2020DevelopmentMoS2, Kravets2019MeasurementsModulators}. On the other hand, focused light on the sample makes it tedious for mechanical variation of the AOI and hence focused-beam SE is usually optimized for a single AOI. This limits the amount of data for acquisition, decreasing the sensitivity of the fit parameters. Another approach to improve lateral resolution is imaging SE which adopts null ellipsometry technique with addition of an objective lens and a two-dimensional detector array to its hardware, achieving a lateral resolution down to 1 micron \cite{Wurstbauer2010ImagingGraphene}. Null ellipsometry limits the imaging SE method to recording single-wavelength (and single AOI) ellipsometric information in a single-shot after mechanical rotation of its polarization components. This makes imaging SE operate with inordinately long measurements times for a spectrally resolved response. In contrast, conventional and focused-beam SE measure broadband responses at a single-AOI in a single-shot and in a short time frame. This fast data acquisition rate of SE holds significant importance for making it a practical tool in industry and research.

To overcome mechanical and technical limitations of ellipsometry, another approach called ``micro-ellipsometry'' has been used \cite{Zhan2002High-resolutionEllipsometer, Zhan2002MicroellipsometerSymmetry, Linke2005QuantitativeInterface, Ye2007Angle-resolvedMicroellipsometry, Tschimwang2010High-spatial-resolutionSymmetry, Otsuki2013BackGeometry, Choi2020Single-shotEllipsometry, Lee2020Co-axialSize, Choi2021CoaxialMeasurement, Tang2021RobustMeasurement, Wang2021ImagingScanning}. Micro-ellipsometry makes use of the NA of an on-axis objective lens for oblique reflection and collection of light. The reflected light with the same AOI is focused on a specific location on the back focal plane, which is imaged by a camera. This approach simplifies collection of angle-resolved ellipsometric information by eliminating the need for mechanical arms, while allowing a micro-scale spot size with its on-axis configuration. Despite its potential advantages, micro-ellipsometry is still not used widely and is not commercially available. One main factor behind this is the lack of an accurate and simple system characterization and calibration. Micro-ellipsometry introduces new system unknowns to the measurement: the AOI values and the instrumental polarization. Theoretical mapping of AOI values on the back focal plane does not take into account the very possible misalignment and errors (especially with manufacturing and assembly errors of high-NA objective lenses) which can have an effect on the AOI locations. This is critical due to high sensitivity of ellipsometry to AOI values. In parallel, the instrumental polarization at each AOI needs to be accurately determined in order to correct for it post-measurement.

Building an accurate, simple, spectroscopic micro-ellipsometer is still an ongoing scientific challenge \cite{Choi2020Single-shotEllipsometry, Lee2020Co-axialSize, Choi2021CoaxialMeasurement, Wang2021ImagingScanning}. Although the importance of accurate AOI characterization has been realized lately \cite{Tang2021RobustMeasurement}, still no method has been proposed for an accurate, simple and complete system characterization to unlock all the advantages of micro-ellipsometry simultaneously, namely, accurate acquisition of angle-resolved spectroscopic ellipsometry data in a single-shot measurement with high lateral resolution.

In this paper we present a single-shot angle-resolved spectroscopic micro-ellipsometer. Its high accuracy is based on our system calibration method that provides a simple, accurate and complete characterization of the optical system. Our instrument is realized with only a few and standard hardware additions to a generic optical microscope, allowing its capabilities to be integrated into standard optical imaging systems in a simple, compact and low-cost manner.

We prove the accuracy of our method by comparing results of complex refractive index and film thickness with a commercial spectroscopic ellipsometer. We demonstrate the high lateral resolution performance by mapping local variations in film thickness and complex refractive index on micro-scale areas. Finally, we show how the combined accuracy and high lateral resolution allows new capabilities by demonstrating measurements of complex refractive index and layer number of tiny flakes of atomically thin exfoliated materials.

\section{Hardware Configuration}

\begin{figure}[H]
\centering
\includegraphics[width=1\textwidth]{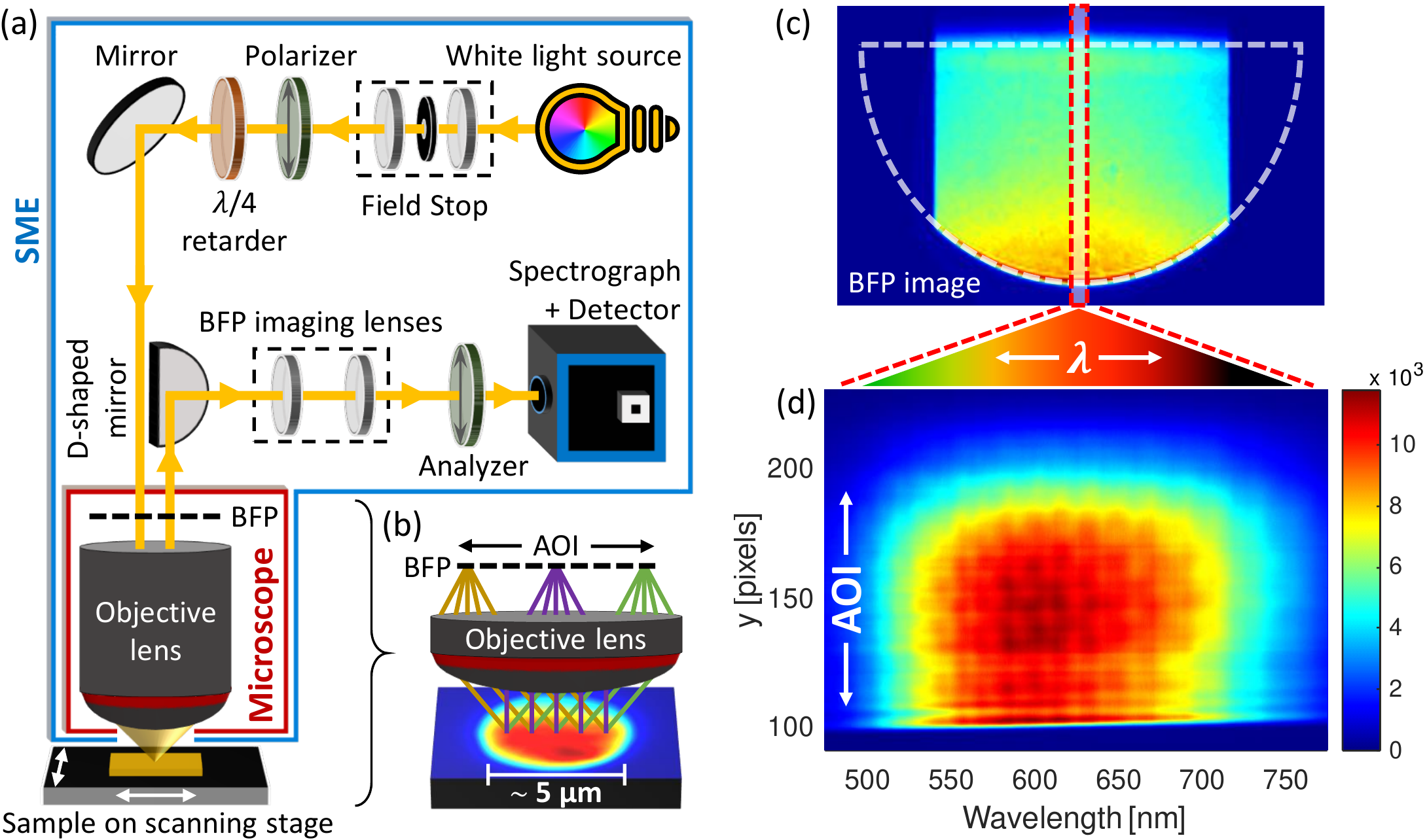}
     \caption{(a) Schematic of the SME system (BFP: Back focal plane). (b) White light reflected from the sample at different AOIs (illustrated by different colors) focusing on different locations on the BFP. The sample is illustrated by a real image of the SME spot with 5 microns diameter. (c) Zero-order image of half-BFP (white dashed lines) cropped by the spectrograph slit from the right and left sides. Narrowing the slit (red dashed lines) and dispersing the image by a diffraction grating yields (d) the first-order image of the BFP center slice as demonstrated at a specific polarization setting. The x-axis is the wavelength and the y-axis is the spatial axis. Each horizontal line of pixels ($y$) corresponds to an AOI value, limited by the NA of the objective lens. AOIs range from around zero ($y=\sim220$) to maximum available by the NA ($y=\sim100$). (The colorbar is of intensity counts.)}
\label{fig:SME_system}
\end{figure}

The Spectroscopic Micro-Ellipsometer (SME) consists a generic optical imaging system (a microscope) with a few added standard optical components and a spectrograph with a two-dimensional detector array, as illustrated in Figure \ref{fig:SME_system}a.

The SME uses a standard, incoherent, quasi-collimated and broadband light source with the desired spectral range. Apart from an external source, the integral broadband illumination of any optical microscope can also be used for SME measurements. A field stop follows the light source, composed of two lenses and a pinhole at their shared focal plane. The pinhole is imaged on the sample with a certain magnification, therefore modulation of the pinhole aperture gives the SME control over its spot size and hence its lateral resolution. Next, a linear polarizer and a quarter-wave ($\lambda$/4) retarder modulate the input polarization by mechanical rotations to be linear or circular, as required for a conventional ellipsometry measurement. The broadband and polarized input illumination is then directed by reflections to pass through the objective lens and reaches the sample with multiple AOIs, confined in a micro-spot. In order to have a large spread of AOIs and a small spot size, high NA objective lenses are preferred. The SME was successfully operated with NA of 0.65 and 0.90. The reflected light from the sample is collected by the same objective lens and is directed by reflections to pass through the back focal plane imaging lenses and another mechanically rotatable linear polarizer (called analyzer), finally reaching the spectrograph with a two-dimensional detector array for spectral analysis. The SME can operate with any standard spectrograph and two-dimensional detector array.

Instead of the D-shaped mirror in Figure \ref{fig:SME_system}a, a non-polarizing beamsplitter can also be used with no effect on the functioning of the SME. D-shaped mirror better preserves the light intensity while refraining from the interference effects that might be caused by some beamsplitters.

All mechanical movement and rotation in the SME is done using computer-controlled motorized stages. The SME system control, data acquisition, measurement and data processing are performed by a custom MATLAB graphical user interface (GUI) code.

\section{Measurement Method}

The SME uses conventional static photometric ellipsometry which measures the spectral intensity of the reflected light at predetermined linear polarization angles, when the input illumination is either linearly or circularly polarized \cite{Azzam1996EllipsometryLight}. These intensity values are functions of wavelength and AOI, and are used to calculate the ellipsometric angles $\Psi$ and $\Delta$ from their mathematical relations to the Stokes parameters \cite{Fujiwara2007SpectroscopicApplications,Roseler1993IRResults}. Using this conventional ellipsometry algorithm, the SME records four intensity images at different polarization settings, as one shown in Figure \ref{fig:SME_system}d, in order to calculate the ellipsometric angles $\Psi$ and $\Delta$.

The back focal plane of the objective lens is the Fourier transform of the spot image on the sample, corresponding to spatially resolved broadband light intensity information of each reflection angle, as illustrated in Figure \ref{fig:SME_system}b. The zero-order image of the back focal plane, seen in Figure \ref{fig:SME_system}c, provides intensity information for multiple AOIs and multiple azimuth angles, with no spectral resolution. Narrowing the spectrograph slit and dispersing the slit image by a diffraction grating results in spectrally resolved first-order intensity distribution of the reflected light for multiple AOIs, as seen in Figure \ref{fig:SME_system}d. Here, the x-axis is the wavelength and the y-axis ($y$) is the vertical pixel values of the detector, corresponding to the range of AOIs provided by the NA of the objective lens.

In order to extract accurate ellipsometric response of the sample from the SME measurement, a complete system characterization and calibration is needed. The AOI values on the back focal plane must be determined with high accuracy, meaning each $y$ value in Figure \ref{fig:SME_system}d must be assigned to its experimental AOI correspondence. In addition, instrumental polarization effects present in the output data must be determined as a function of wavelength and AOI. The proposed method in this work measures these system unknowns with high accuracy by using only experimental ellipsometric calibration measurements, as explained in the next section. The direct measurement results from the SME can then be calibrated with this system data for sample-only ellipsometric information.

\section{Characterization \& Calibration Method}

The SME system characterization and calibration method \cite{Kenaz2022SystemEllipsometry} does not require any prior theoretical knowledge on the setup. This fully experimental technique allows for highly accurate ellipsometric measurements in a generic optical imaging configuration.

The first step of the method is measuring the complex refractive indices for two different reference samples of evaporated or sputtered optically opaque noble metals (we use gold and platinum) by a commercial spectroscopic ellipsometer. From these measurements, their $\Psi$ and $\Delta$ values can be extracted at any AOI. Then the same materials are measured with the SME and the $\Psi$ and $\Delta$ values are obtained, with yet to be resolved AOI values and included instrumental polarization. These measurements of the reference samples by the commercial and SME instruments are the input data needed for the complete system characterization. 

Next, a single AOI ($y$) data is selected from the SME measurements for its characterization and calibration. A range of possible AOI values are assigned to this data, and at each assignment the corresponding instrumental polarization candidate is calculated with the input of extracted $\Psi$ and $\Delta$ from the commercial ellipsometer measurement at the assigned AOI. The system is represented collectively as a virtual sample and the instrumental polarization is defined by ellipsometric parameters $\Psi_{ins}$ and $\Delta_{ins}$. This calculation of instrumental polarization candidates for a selected $y$ data is performed in both reference sample measurements separately.

\begin{figure}[H]
\centering
\includegraphics[width=0.8\textwidth]{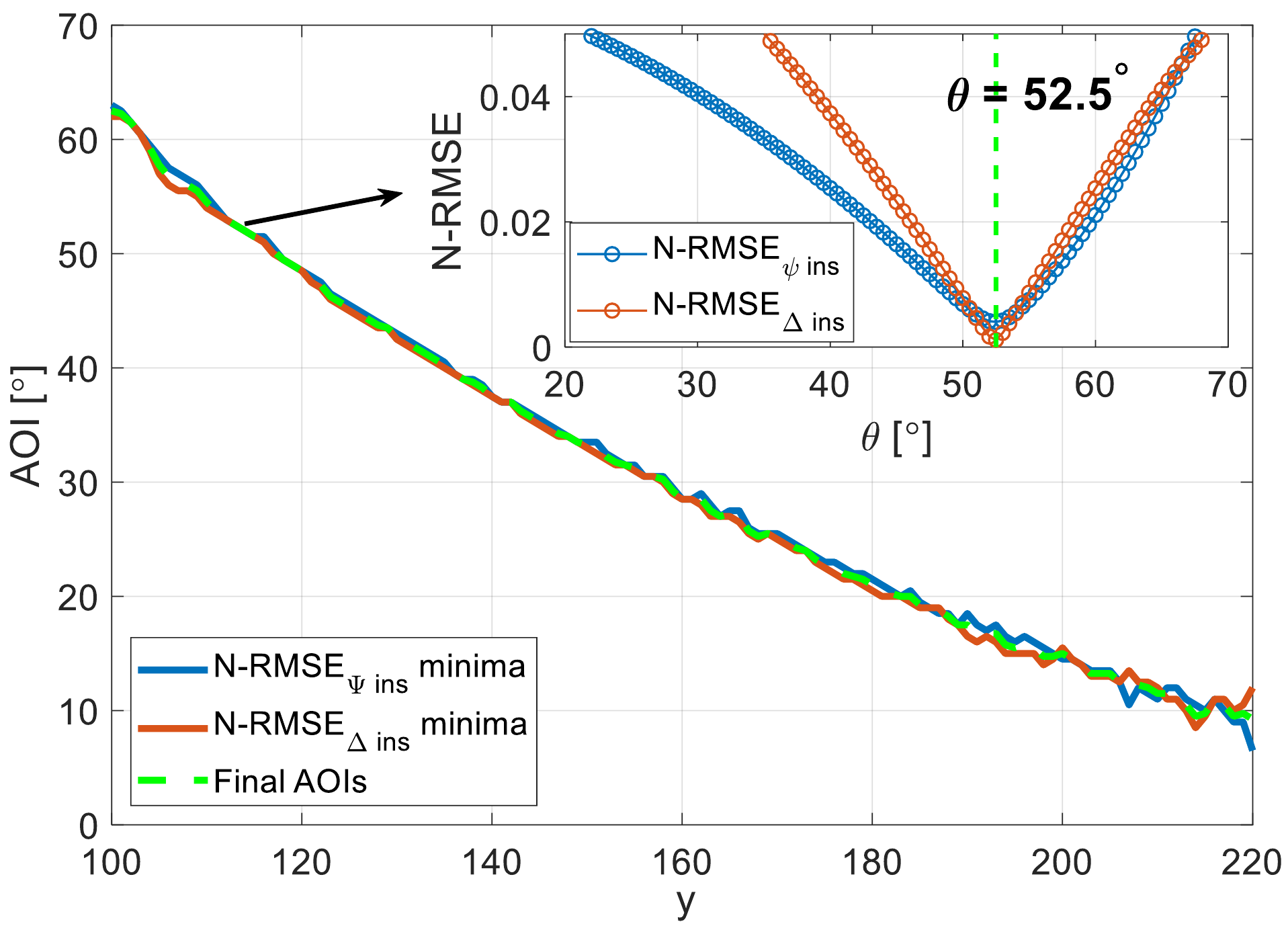}
     \caption{The AOI characterization plot showing N-RMSE$_{\Psi ins}$ and N-RMSE$_{\Delta ins}$ minima positions versus the radial position on the back focal plane ($y$). The average of (solid blue and orange) minima positions is taken as the final AOIs (dashed green). Inset: N-RMSE$_{\Psi ins}$ and N-RMSE$_{\Delta ins}$ comparison values for sequential $\theta$ assignments at $y$=113 position, pointing to common minima at $\theta=52.5^{\circ}$.}
\label{fig:AOI_calibration}
\end{figure}

The characterization method is based on the assumption that the properties of a stable optical imaging system are constant and independent of the measured sample. Therefore, the instrumental polarization candidates $\Psi_{ins}$ and $\Delta_{ins}$ are compared between the two reference measurements (by normalized root-mean-square error, N-RMSE) and plotted as a function of assigned AOIs, $\theta$. As shown in the inset of Figure \ref{fig:AOI_calibration}, similar v-shaped curves are obtained for $\Psi_{ins}$ and $\Delta_{ins}$ comparisons with their minima pointing to the same $\theta$ value. This indicates the existence of a single AOI value where the instrumental polarization in the two reference measurements overlap, proving the sample-independent and unique system properties. This singular intersection point reveals the experimental AOI value and its corresponding instrumental polarization simultaneously.

Repeating this procedure for all $y$ positions reveals the complete AOI map of the back focal plane as seen in Figure \ref{fig:AOI_calibration}, together with the corresponding instrumental polarization at each AOI. In cases of slight discrepancy between the minima positions, the average is taken as the final AOI.

The detailed characterization and calibration procedure can be found in Supplementary Information (SI) section \ref{detailedmethod}.

Once characterized for a stable system, the extracted system data can be stored to calibrate future measurements for accurate ellipsometric results.

Since the method consists a general AOI and polarization characterization for an optical imaging system, it can also be used with other schemes of micro-ellipsometry or with any other measurement where AOI and instrumental polarization information are critical. In case azimuthal angles of reflection are of interest, the method can be implemented on a monochromatic basis without a spectrograph. Although these schemes are not experimented yet, they are theoretically feasible.

\section{Results: Performance \& Capabilities}

Following calibration, the SME is capable of recording angularly and spectrally resolved ellipsometric data ($\Psi$ and $\Delta$) from a spot area of a few microns in a few seconds. The single-shot measurement time around 10 seconds is dependent on system hardware and is mostly limited by the mechanical rotation of polarization elements and exposure time at each intensity recording. The high amount of spectrally and angularly resolved data acquired in this short time frame practically gives the SME better sensitivity and accuracy in data fitting.

First, the accuracy of the SME is tested by comparing results of both complex refractive index and film thickness to a commercial spectroscopic ellipsometer. Then, its high lateral resolution is demonstrated by mapping film thickness and complex refractive index variations in micro-scale areas, with a spot size down to 2 microns. Finally, the high accuracy and lateral resolution of the SME is used to measure the optical properties and thickness of micro-scale flakes of exfoliated atomically thin materials, and results are compared to previous works in the literature.

For modelling and fitting, WVASE\textsuperscript{\tiny\textregistered} and CompleteEASE\textsuperscript{\tiny\textregistered} ellipsometry data analysis software from the J.A. Woollam Co., Inc. are used.

\subsection{Complex refractive index}

\begin{figure}[H]
\centering
\includegraphics[width=0.75\textwidth]{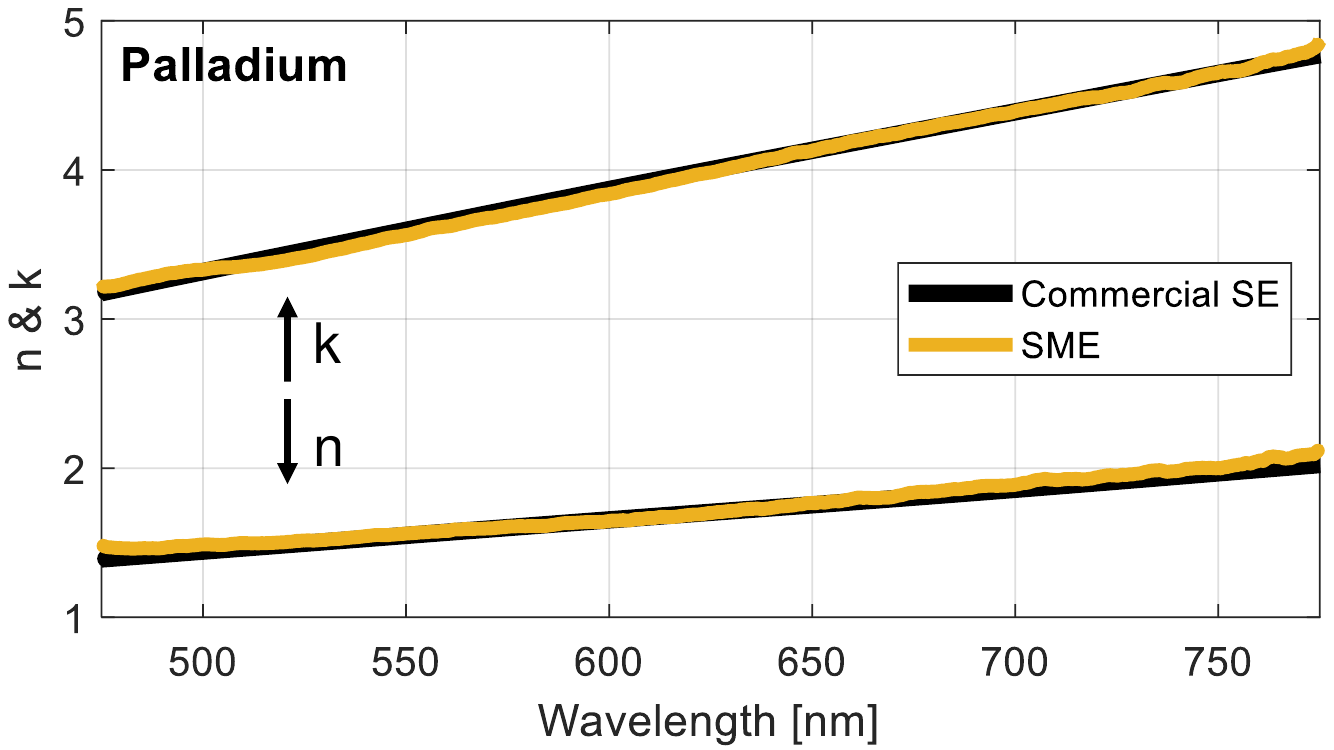}
     \caption{Complex refractive index (n \& k) of palladium, measured by the SME (5 $\mu m$ spot diameter) and by a commercial ellipsometer (3 mm $\times$ 9 mm spot size).}
\label{fig:palladium_nk}
\end{figure}

The calibrated SME is used for complex refractive index measurement of a plain palladium surface and result is compared with a commercial spectroscopic ellipsometer (J.A. Woollam alpha-SE) having a 3 mm $\times$ 9 mm spot size, as seen in Figure \ref{fig:palladium_nk}. A single-shot measurement is performed by the SME with a spot diameter of 5 microns and results from multiple AOIs are averaged. For an explicit demonstration of the instrument-only result, no data fitting or smoothing is applied to the SME result. A good agreement is demonstrated between the two instruments, proving the accuracy of the SME in measuring complex refractive indices.

It is important to note that slight variations in results of the two instruments are expected due to six orders-of-magnitude difference in the measurement spot areas.

\subsection{Film thickness} 

For demonstration of film thickness measurement accuracy of the SME, different thicknesses of SiO$_2$ films with nominal values of 285 nm, 90 nm, 60 nm and 25 nm deposited on silicon substrates are used. The calibrated SME single-shot data from nominal 285 nm of SiO$_2$ on Si can be seen in SI Figure \ref{fig:285data}.

\begin{figure}[H]
\centering
\includegraphics[width=0.9\textwidth]{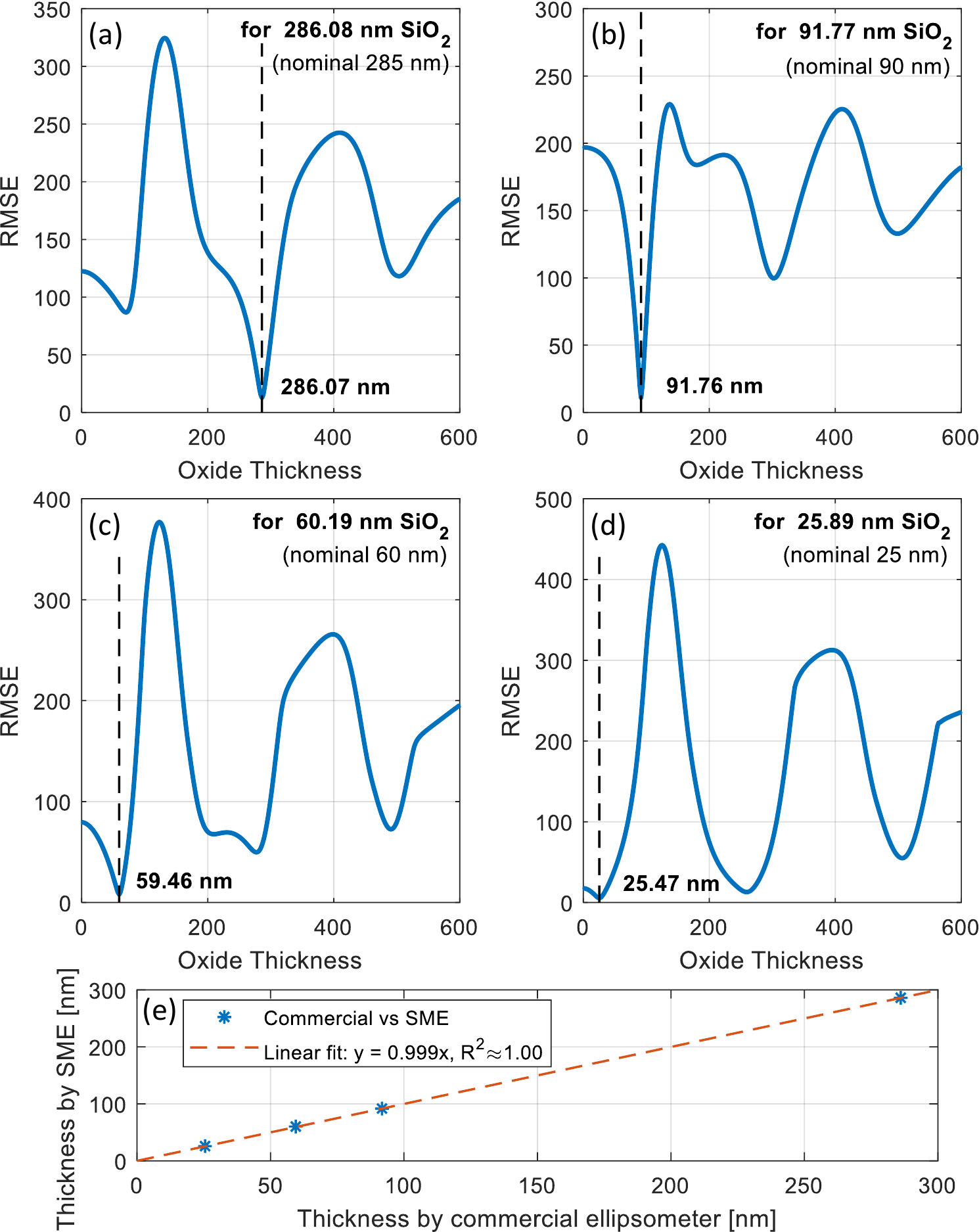}
     \caption{(a-d) The SME parameter uniqueness plots for SiO$_2$/Si samples with commercially measured oxide layer thicknesses written on the top-right corner of each plot. The SME results are written near the well-defined global minima, showing good agreement with the commercial results. (e) Plot of SiO$_2$ thicknesses measured by the commercial ellipsometer versus the SME. Accuracy of the SME is demonstrated with the high agreement to the linear fit (R$^2$: goodness of fit).}
\label{fig:uniqueness}
\end{figure}

In ellipsometry data analysis, parameter uniqueness plots show the variation of the error (in RMSE) between the model and the data when the model is scanned for a selected fit parameter. The parameter uniqueness plots of nominal 285 nm, 90 nm, 60 nm and 25 nm SiO$_2$ measurements by the SME can be seen in Figure \ref{fig:uniqueness}a-d when the models are scanned for SiO$_2$ thickness values from zero to 600 nm. In all measurements, well-defined global minima show the SME results (written near the minima). The same samples are also measured by a commercial spectroscopic ellipsometer (J.A. Woollam alpha-SE) having a spot size of 3 mm $\times$ 9 mm. Very similar results to the SME results are obtained (written at the top-right corner of each plot), up to inherent slight variations due to the significant difference between the spot sizes. In both SME and commercial data analysis, the same models and material parameters are used for a fair comparison. Figure \ref{fig:uniqueness}e plots and fits the SiO$_2$ thickness results measured by the commercial ellipsometer versus the SME and obtains high linear agreement, confirming the accuracy of the SME.

In order to confirm the repeatability of the measurement devices, each measurement by the commercial spectroscopic ellipsometer and the SME are repeated multiple times and similar results are received consistently.

\subsection{Spatial mapping}

\begin{figure}[H]
\centering
\includegraphics[width=0.95\textwidth]{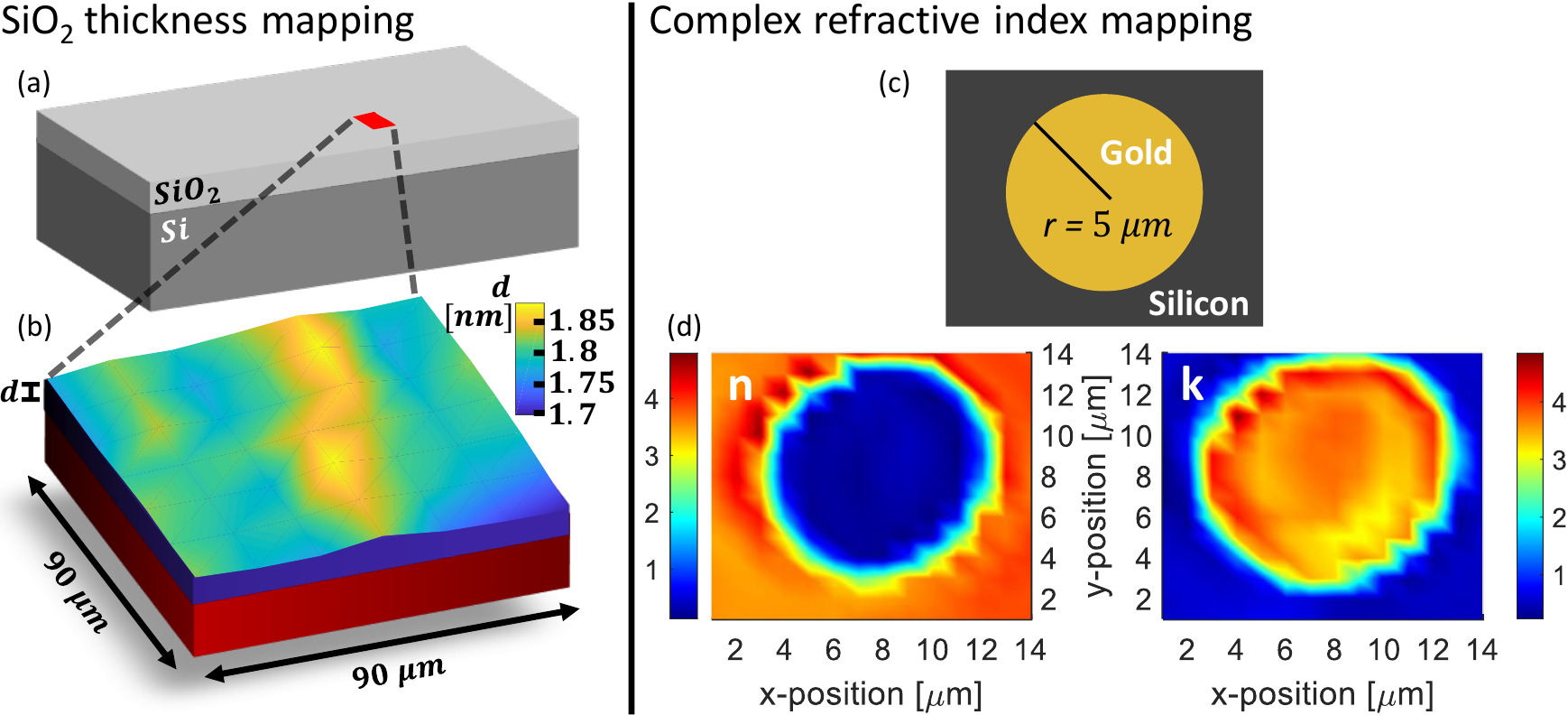}
     \caption{(a) Illustration of a native SiO$_2$ layer on Si, (b) mapped by the SME for its oxide thickness variations in an area of 90 $\mu m$ $\times$ 90 $\mu m$ with a spot diameter of 7 $\mu m$ and a step size of 15 $\mu m$. (c) Illustration of a gold disc with radius of 5 $\mu m$ on a silicon substrate, (d) mapped by the SME for its optical constants (n \& k) at $\lambda$= 632 nm with a spot diameter of 2 $\mu m$ and a step size of 1 $\mu m$.}
\label{fig:mappings}
\end{figure}

The spatial mapping capability of the SME is demonstrated by thickness and refractive index maps of micro-scale areas with a high lateral resolution. The sample under measurement sits on a two-axis computer-controlled stage with micrometer movement resolution (as illustrated in Figure \ref{fig:SME_system}a), providing the SME with spatial mapping capability. The local thickness profile of a native oxide layer on a silicon wafer is mapped in an area of 90 $\mu m$ $\times$ 90 $\mu m$ and compared to the thickness result from a commercial spectroscopic ellipsometer (J.A. Woollam alpha-SE) with 3 mm $\times$ 9 mm spot size. For this measurement, a larger spot diameter (7 $\mu m$) and a large step size (15 $\mu m$) are used to map the expected slight variations in oxide thickness on a wide area. The sample illustration and thickness profile result can be seen in Figure \ref{fig:mappings}a-b. The average thickness value of the 49 measurement points is 1.8 nm, which is identical to the result by the commercial ellipsometer.

To further demonstrate the SME's high lateral resolution capability, a micro-structured area of 14 $\mu m$ $\times$ 14 $\mu m$ is spatially scanned with a spot size of 2 $\mu m$ and a step size of 1 $\mu m$. The vicinity of a gold disc with a radius of 5 $\mu m$ on a silicon substrate is mapped for local variations in optical constants. The structure illustration and the complex refractive index components $n$ and $k$ plotted at 632 nm wavelength are shown in Figure \ref{fig:mappings}c-d. Here the complex refractive indices are calculated directly from the measured ellipsometric data, as stated in SI Equation \ref{eq:N1}.

\subsection{The SME on exfoliated two-dimensional materials}
\label{2Dsection}

The combined accuracy and high lateral resolution of the SME allows measurements of complex refractive index and layer number of tiny flakes of atomically thin exfoliated materials. Single and few-layer two-dimensional materials are a widely studied topic due to being promising candidates for future photonic and electronic devices \cite{Choi2016Two-dimensionalPhotonics}. Mechanical exfoliation of these materials is simple, low-cost and results in single and few-layer flakes with very high purity. However, ellipsometric characterization of these materials is still an ongoing scientific challenge since single-shot spectroscopic ellipsometers cannot resolve these materials easily because of their micro-scale lateral sizes mostly up to around 20 microns.

For testing the measurement sensitivity of the proposed method with the thinnest materials available and addressing a modern scientific challenge, SME measurements are made on mechanically exfoliated molybdenum disulfide (MoS$_2$) and graphene, and the results are compared with previous works in the literature.

\begin{figure}[H]
\centering
\includegraphics[width=0.95\textwidth]{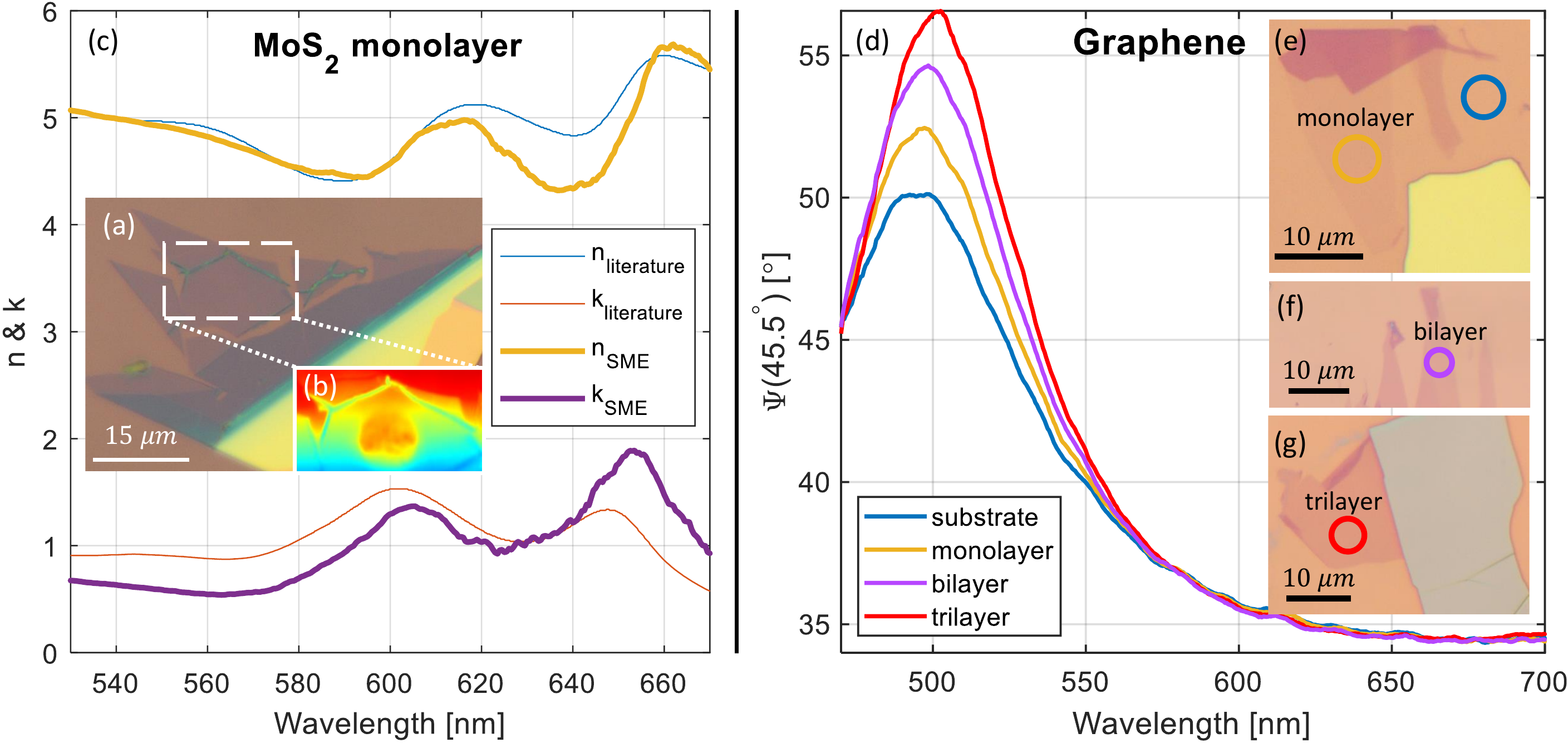}
     \caption{(a) Optical microscope image of exfoliated MoS$_2$ mono- and multi-layers on SiO$_2$/Si substrate. (b) The SME detector image of the marked area in (a), with the SME measurement spot diameter of around 5 microns. (c) Wavelength-by-wavelength calculated complex refractive index (n \& k) of the MoS$_2$ monolayer with comparison to a fitted micro-reflectance result from the literature \cite{Hsu2019Thickness-DependentWSe2}. (d) The SME $\Psi$ values at $45.5^{\circ}$ AOI from marked areas on (e) SiO$_2$/Si substrate, monolayer, (f) bilayer and (g) trilayer graphene flakes, showing distinctive variations around 500 nm wavelength, as shown similarly in the literature \cite{Kravets2010SpectroscopicAbsorption, Funke2017Spectroscopic2D-materials}.}
\label{fig:mos2_graphene}
\end{figure}

The SME is used for complex refractive index measurement of a mechanically exfoliated MoS$_2$ monolayer (confirmed by Raman spectroscopy, SI Section \ref{ramansection}) on nominal 285 nm SiO$_2$ on Si substrate, as seen in Figure \ref{fig:mos2_graphene}a-c. For this measurement, first the SiO$_2$ thickness is measured by the SME in proximity of the MoS$_2$ monolayer. Then a measurement on the monolayer is performed. The structure is modeled from top to bottom as Air/MoS$_2$/SiO$_2$/Si, and the previously measured SiO$_2$ thickness value is entered into the model with the assumption that it remains the same under the monolayer. A thickness of 0.63 nm is assigned to the MoS$_2$ layer and its complex refractive index is calculated by wavelength-by-wavelength (point-by-point) fit in the spectral region of its A and B excitons around 1.90 eV (652.5 nm) and 2.05 eV (604.8 nm) respectively \cite{Funke2016ImagingMoS2}. Wavelength-by-wavelength fit directly calculates the optical constants at each spectral point independent of the neighboring spectra in order to find the best match to the experimental data. By this fitting method, the aim is to demonstrate explicitly the SME's sensitivity to the optical properties of a single layer of molecules. The wavelength range with the highest relative signal-to-noise ratio in the measurement was selected for obtaining a less noisy result. The SME result was compared to a micro-reflectance measurement on a mechanically exfoliated MoS$_2$ monolayer \cite{Hsu2019Thickness-DependentWSe2} with a good agreement, proving the sensitivity and accuracy of the SME. Some discrepancy in the results with the literature was expected due to difference in samples and methods. As far as known to us, this is the first single-shot spectroscopic ellipsometry measurement on a mechanically exfoliated transition-metal dichalcogenide monolayer. Here the physics of this result is not further analyzed as it is out of the scope of this paper. 

Next, the SME is used for measurements on exfoliated graphene, a single atomic layer material \cite{Novoselov2004ElectricFilms}. Spectroscopic ellipsometry measurements were performed on exfoliated monolayer, bilayer and trilayer graphene flakes and it was shown to be an effective tool for quantitatively distinguishing between different layer numbers of graphene flakes \cite{Kravets2010SpectroscopicAbsorption}. In that work, the considerably large-sized graphene flakes of around 100 microns allowed measurements by a commercial focused-beam spectroscopic ellipsometer. Clearly distinguishable $\Psi$ values at $45^{\circ}$ AOI for the substrate ($\sim$300 $\mu m$ SiO$_2$ on Si), monolayer, bilayer and trilayer around the wavelength of 500 nm was reported, which can be used as an indicator for the layer number of graphene flakes. Similar measurements are performed by the SME on mechanically exfoliated monolayer, bilayer, trilayer graphene flakes (confirmed by Raman spectroscopy, SI Section \ref{ramansection}) on a substrate of 286.12 $\mu m$ SiO$_2$ on Si, as shown in Figure \ref{fig:mos2_graphene}e,f,g respectively. Here the sizes of graphene flakes are all less than 20 microns (as commonly obtained by mechanical exfoliation), making them not suitable for measurement by focused-beam spectroscopic ellipsometers. Distinguishable $\Psi$ values are observed on different layer numbers of graphene flakes and the substrate in a very similar manner to the reference \cite{Kravets2010SpectroscopicAbsorption}, as seen in Figure \ref{fig:mos2_graphene}d, proving the sensitivity of the SME to single atomic layers.

The SME performance on small-area exfoliated two-dimensional materials demonstrate the strength of the proposed method and its high potential for future applications when its ellipsometric accuracy and sensitivity are combined with its high lateral resolution. As a simple and affordable system which can be integrated into generic optical microscopes, the SME holds the potential to be the new practical tool for characterization and layer number identification of two-dimensional materials (which will be discussed in a future publication).

\section{Summary and Conclusions}

We have demonstrated a fast, mapping, spectroscopic micro-ellipsometer with a lateral resolution down to 2 microns. In a single-shot, the SME can measure the optical properties or thickness of thin samples on a wide wavelength range with fine spectral and angular resolution, limited by the choice of system hardware.

We showed that the accuracy of a microscope ellipsometer can be significantly improved using a full characterization and calibration method that effectively eliminates uncertainties in angle and system response, allowing performance at a level of standard ellipsometers but with a high lateral resolution and high data acquisition rate. 

We demonstrated the accurate performance of the SME by benchmarking it against standard commercial ellipsometers, with excellent agreement in both complex refractive index and thickness measurements. Then we have used the combined power of accuracy and high lateral resolution to demonstrate highly sensitive measurements performed on exfoliated two-dimensional materials.

The multiple-angle and spectroscopic data the SME can collect in a single-shot of few seconds makes it fast in ellipsometric data collection. This high amount of data available in a short time for analysis practically increases the sensitivity of the fit parameters and yields high accuracy in the results. Currently, the measurement time is mostly limited by mechanical rotation of optics for polarization modulation and total exposure time. Polarization modulation time can be significantly decreased by eliminating mechanical rotations and using electro-optic or photoelastic modulators instead. The exposure time can also be decreased by using a higher intensity light source or a more sensitive detector.

The SME includes only standard optical components and can easily be integrated into new or existing generic optical microscopes as an add-on unit. The strength of the SME comes from its system characterization and calibration method, allowing practical implementation of spectroscopic micro-ellipsometry.

Compared to focused-beam spectroscopic ellipsometers with relatively larger spot sizes, the SME boosts the lateral resolution by an order-of-magnitude without compromising ellipsometric accuracy. As of data acquisition rate, the SME collects spectrally resolved ellipsometric data from multiple AOIs in a single-shot of a few seconds  (as seen in SI Figure \ref{fig:285data}), outperforming both focused-beam and imaging spectroscopic ellipsometers by at least one and three orders-of-magnitude, respectively. Furthermore, the SME achieves these advantages in a compact, affordable, microscope-integrated design.

\section{Acknowledgements}

We want to thank to Dr. Pradheesh Ramachandran and Prof. Hadar Steinberg for preparation of 2D materials samples, and to Israel Innovation Authority for its financial support.

\section{Disclosures}

The authors declare no conflicts of interest.

\section{Data availability}

Data underlying the results presented in this paper are not publicly available at this time but may be obtained from the authors upon reasonable request.

\clearpage
\bibliography{references}
\bibliographystyle{ieeetr}

\newpage

\supplementaryinformation

\section{Characterization \& Calibration Procedure}
\label{detailedmethod}

In the first step of our calibration method \cite{Kenaz2022SystemEllipsometry}, two different reference materials are measured for their complex refractive indices by a commercial spectroscopic ellipsometer (J.A. Woollam alpha-SE). Ellipsometry measures the complex reflectance ratio ($\rho$) of a system, which may be written in terms of the amplitude ratio component $\Psi$ and the phase difference component $\Delta$, as seen in Equation \ref{eq:rho}. $\Psi$ and $\Delta$ are, by definition, functions of wavelength ($\lambda$) and angle of incidence (AOI).

\begin{equation}
\rho=\tan\Psi\:e^{i \Delta}
\label{eq:rho}
\end{equation}

Single layer, optically opaque, homogeneous and isotropic materials (e.g. noble metals) allow a direct mathematical transformation between their complex refractive indices and their ellipsometric parameters, as seen in Equation \ref{eq:N1}:

\begin{equation}
N_1=n+ik=N_{0}\tan\theta\left[1-\frac{4\rho}{(1+\rho)^2}\sin^2\theta\right]^{\frac{1}{2}}
\label{eq:N1}
\end{equation}

where $N_{1}$ is the complex refractive index of the material under investigation, $N_{0}$ is the complex refractive index of the incident medium (which is mostly air with $N_{0}$=1), $\theta$ is the AOI value, and $\rho$ is the complex reflectance ratio.

We use gold and platinum as reference materials, deposited either by evaporation or sputtering. After their complex refractive index measurements by the commercial spectroscopic ellipsometer, Equation \ref{eq:N1} allows calculation of $\Psi_{ref_{1,2}}$ and $\Delta_{ref_{1,2}}$ values for the two reference materials at any AOI. The subscripts 1 and 2 represent the gold and platinum measurements, respectively.

Then the same materials are measured by the SME. The raw experimental results $\Psi_{exp_{1,2}}$ and $\Delta_{exp_{1,2}}$ are calculated directly from the intensity values at different polarization settings, by the help of Stokes parameters \cite{Fujiwara2007SpectroscopicApplications,Roseler1993IRResults}. However, they are not representative of the reference samples because of the included unknown instrumental polarization, and cannot be used for extracting any useful physical information because of the unknown AOI values.

Our method assumes the properties of a stable optical system to be constant and independent of the measured sample, hence presumes the above mentioned system unknowns to be identical in the two reference measurements performed by the SME. This sample-independent system information requires the system components and alignment to be identical - and not necessarily ideal - for all performed measurements. The system is treated collectively as a virtual sample and the wavelength-and-AOI-dependent instrumental polarization is represented by $\Psi_{ins}$ and $\Delta_{ins}$.

The sample and instrument responses add up on the polarization state in the final reading from the SME. Therefore, the measured raw complex reflectance ratio $\rho_{exp}$ by the SME can be written as a function of complex reflectance ratios of the reference sample ($\rho_{ref}$) and instrument ($\rho_{ins}$) in the two reference measurements, as:

\begin{equation}
\rho_{exp_{1,2}}=\rho_{ref_{1,2}}\cdot\rho_{ins}=\tan(\Psi_{exp_{1,2}})\:e^{i\,\Delta_{exp_{1,2}}}
\label{eq:rho_exp}
\end{equation}

This allows writing the instrumental parameters as functions of the raw experimental and the reference parameters:

\begin{equation}
\Psi_{ins_{1,2}}=\tan^{-1}\left(\frac{\tan(\Psi_{exp_{1,2}})}{\tan(\Psi_{ref_{1,2}})}\right)
\label{eq:Psi_ins}
\end{equation}

\begin{equation}
\Delta_{ins_{1,2}}=\Delta_{exp_{1,2}}-\Delta_{ref_{1,2}}
\label{eq:delta_ins}
\end{equation}

Here, the instrumental parameters are defined separately for the two reference measurements (as $\Psi_{ins_{1,2}}$ and $\Delta_{ins_{1,2}}$) which are both identical to instrumental polarization for the system ($\Psi_{ins}$ and $\Delta_{ins}$) independent of the measured sample. This representation is temporarily used for the purpose of characterization. Our method is based on searching for the condition of this unique and system-specific instrumental polarization.

For a single $y$ position (as in Figure \ref{fig:SME_system}d), the raw experimental parameters ($\Psi_{exp_{1,2}}$, $\Delta_{exp_{1,2}}$) are calculated in the two reference measurements. Next, this raw experimental data is assigned to a range of $\theta$ (AOI) values with a desired angle resolution (depending on the limits of the optical system). At each $\theta$ assignment, the instrumental parameters candidates ($\Psi_{ins_{1,2}}$ and $\Delta_{ins_{1,2}}$) are calculated via the input of calculated reference parameters ($\Psi_{ref_{1,2}}$, $\Delta_{ref_{1,2}}$) at that assigned $\theta$ into Equations \ref{eq:Psi_ins} and \ref{eq:delta_ins}.

Then, the instrumental polarization candidates are compared between the two reference measurements ($\Psi_{ins_{1}}$ versus $\Psi_{ins_{2}}$, and $\Delta_{ins_{1}}$ versus $\Delta_{ins_{2}}$) and plotted as a function of assigned $\theta$ values. For quantitative comparison, normalized root-mean-square-error (N-RMSE) is used, defined as:

\begin{align}
\text{N-RMSE}_{A}(\theta)&=\sqrt{\frac{1}{N}\sum_{p=1}^{N}\left[\frac{(A_1(\lambda_{p},\theta)-A_2(\lambda_{p},\theta))}{\frac{1}{2}(A_1(\lambda_{p},\theta)+A_2(\lambda_{p},\theta))}\right]^{2}}
\label{eq:RMSE_equation}
\end{align}

where $A$ is either $\Psi_{ins}$ or $\Delta_{ins}$, subscripts 1 and 2 are the gold and platinum reference measurements respectively, $p$ is the sequence number of a single wavelength in the spectral range and $N$ is the total number of wavelength points.

Plotting N-RMSE$_{\Psi_{ins}}$ and N-RMSE$_{\Delta_{ins}}$ as functions of assigned $\theta$ values results in well-defined common minima at a specific $\theta$, as seen in the inset of Figure \ref{fig:AOI_calibration}, pointing to where the instrumental polarization candidates from two different reference measurements are nearly identical. This overlap of instrumental response at a single $\theta$ proves the assumption of sample-independent system information, revealing the experimental AOI value and the corresponding instrumental polarization at the selected experimental data position ($y$) simultaneously.

Repeating this process for the desired range of $y$ positions characterizes the AOIs and their corresponding instrumental polarization. In cases where slight discrepancies between the $\theta$ values at minima of N-RMSE$_{\Psi ins}$ and N-RMSE$_{\Delta ins}$ are observed (probably due to some random errors), their average is taken to conclude the final AOI values, represented by the green dashed line in Figure \ref{fig:AOI_calibration}. In parallel, the matching instrumental polarization from the two reference measurements are also averaged at the final AOI value for a unique system response.

As seen in Figure \ref{fig:AOI_calibration}, the maximum AOI available in the system is measured to be $62.5^{\circ}$, which is in good agreement with the used NA of 0.9, theoretically allowing until around $64^{\circ}$. As the AOI gets smaller, the minima agreement gets noisier in an expected manner with the decreasing degree of polarization variation.

After the AOI and corresponding instrumental polarization characterization, the calibrated results can be calculated by rearranging Equations \ref{eq:Psi_ins} and \ref{eq:delta_ins} and replacing reference parameters with sample results, as seen in Equations \ref{eq:psi_sample} and \ref{eq:delta_sample}:

\begin{equation}
\Psi_{sample}=\tan^{-1}\left(\frac{\tan(\Psi_{exp})}{\tan(\Psi_{ins})}\right)
\label{eq:psi_sample}
\end{equation}

\begin{equation}
\Delta_{sample}=\Delta_{exp}-\Delta_{ins}
\label{eq:delta_sample}
\end{equation}

\begin{figure}[H]
\centering
\includegraphics[width=0.9\textwidth]{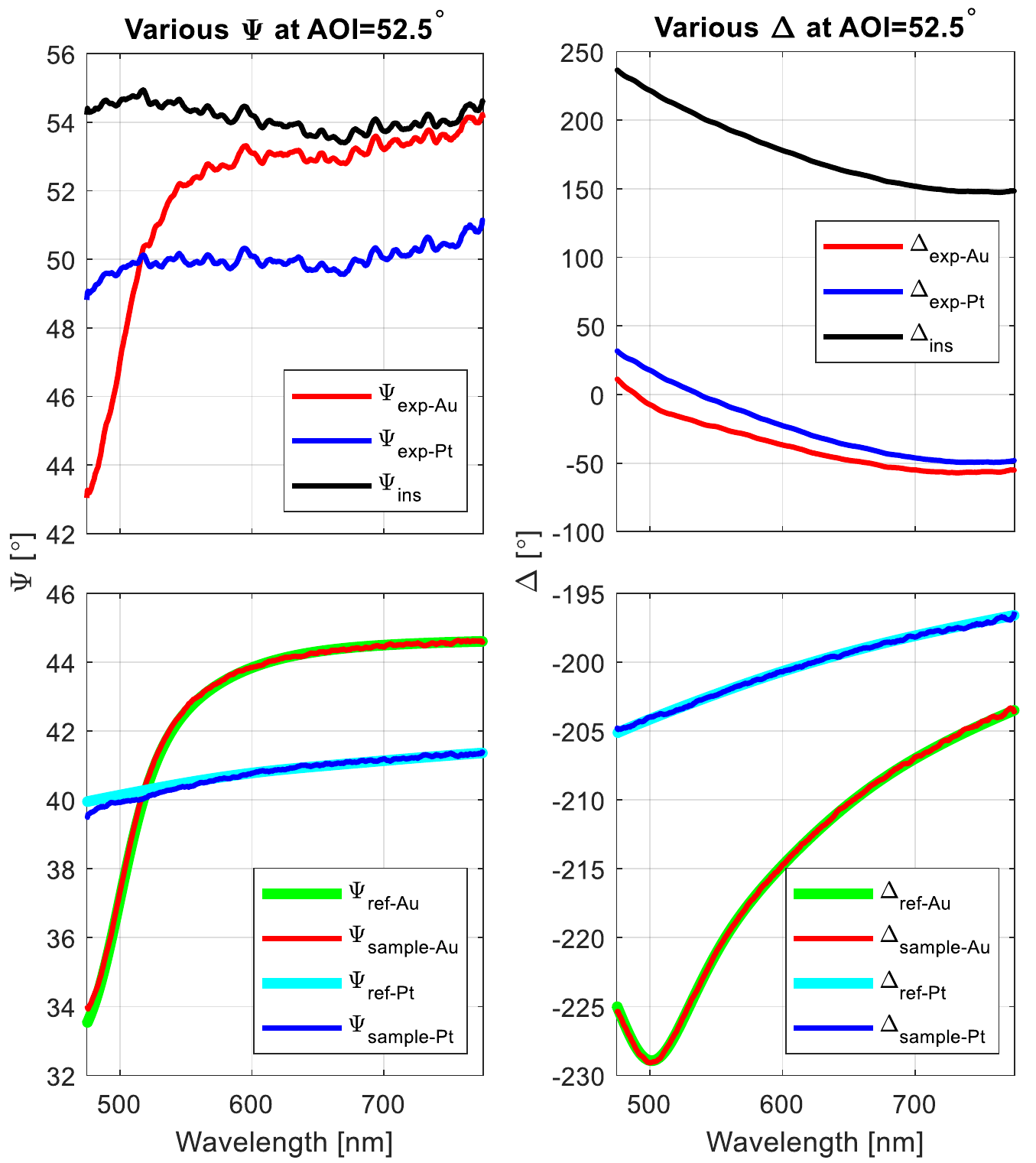}
     \caption{Various $\Psi$ and $\Delta$ data for gold (Au) and platinum (Pt) measurements at $y$=113 $\rightarrow$ AOI=$52.5^{\circ}$. (exp: raw experimental, ins: instrumental, ref: reference)}
\label{fig:SMEdata}
\end{figure}

The instrumental, raw experimental and reference $\Psi$ and $\Delta$ values at $y$=113 $\rightarrow$ AOI=$52.5^{\circ}$ (as shown in Figure \ref{fig:AOI_calibration} inset) are plotted in Figure \ref{fig:SMEdata}. The $\Psi_{sample-Au,Pt}$ and $\Delta_{sample-Au,Pt}$ for reference gold (Au) and platinum (Pt) samples are the reconstructed values by Equations \ref{eq:psi_sample} and \ref{eq:delta_sample}, with the input of raw experimental results ($\Psi_{exp-Au,Pt}$ and $\Delta_{exp-Au,Pt}$) and the extracted system parameters ($\Psi_{ins}$, $\Delta_{ins}$). These reconstructed parameters are then compared to the reference parameters from the commercial ellipsometer at the extracted AOI value. The achieved agreement visually demonstrates that only a unique AOI and instrumental polarization at a single $y$ position will calibrate the raw experimental results of both materials to be in a good agreement with the calculated reference results.

\begin{figure}[H]
\centering
\includegraphics[width=0.9\textwidth]{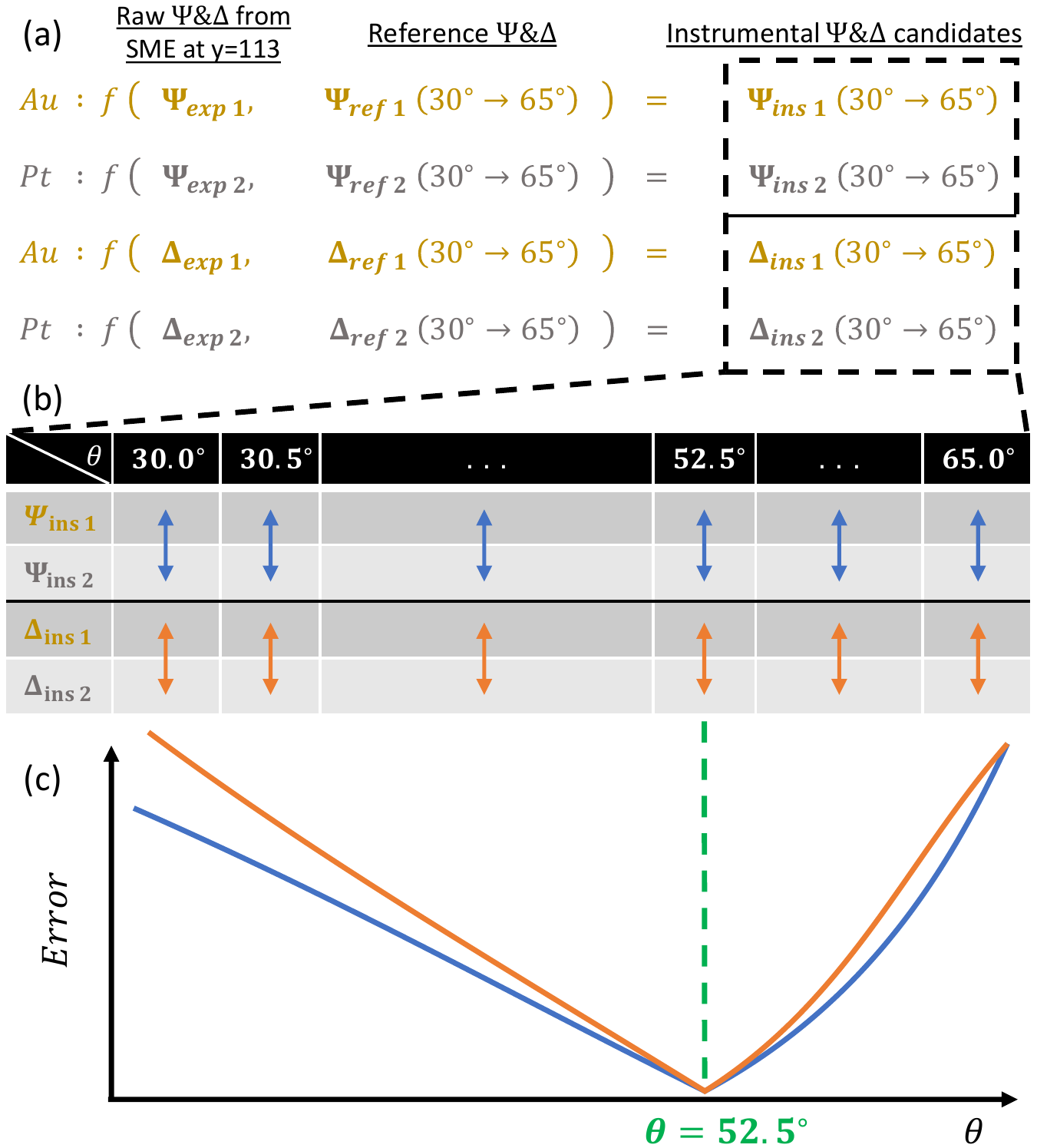}
     \caption{Illustrative summary of the system characterization algorithm at $y$=113, with reference to Figure \ref{fig:AOI_calibration} inset. (a) At $y$=113 position of gold (Au, 1) and platinum (Pt, 2) SME measurements, the instrumental $\Psi$\&$\Delta$ candidates are calculated as a function of the raw experimental $\Psi$\&$\Delta$ and reference $\Psi$\&$\Delta$ at the selected range of $\theta$ (AOI) values, illustrated here for $30^{\circ}$ to $65^{\circ}$ with a resolution of $0.5^{\circ}$. (b) The instrumental $\Psi$\&$\Delta$ candidates are compared quantitatively between the two reference measurements and (c) plotted for the error as a function of assigned sequential $\theta$ values, resulting in common minima at $\theta=52.5^{\circ}$ for both instrumental $\Psi$ (blue curve) and $\Delta$ (orange curve) candidates comparisons.}
\label{fig:method}
\end{figure}

An illustrative summary for single $y \rightarrow$ AOI characterization is demonstrated in Figure \ref{fig:method}.

\section{The SME single-shot ellipsometric data}

\begin{figure}[H]
\centering
\includegraphics[width=0.9\textwidth]{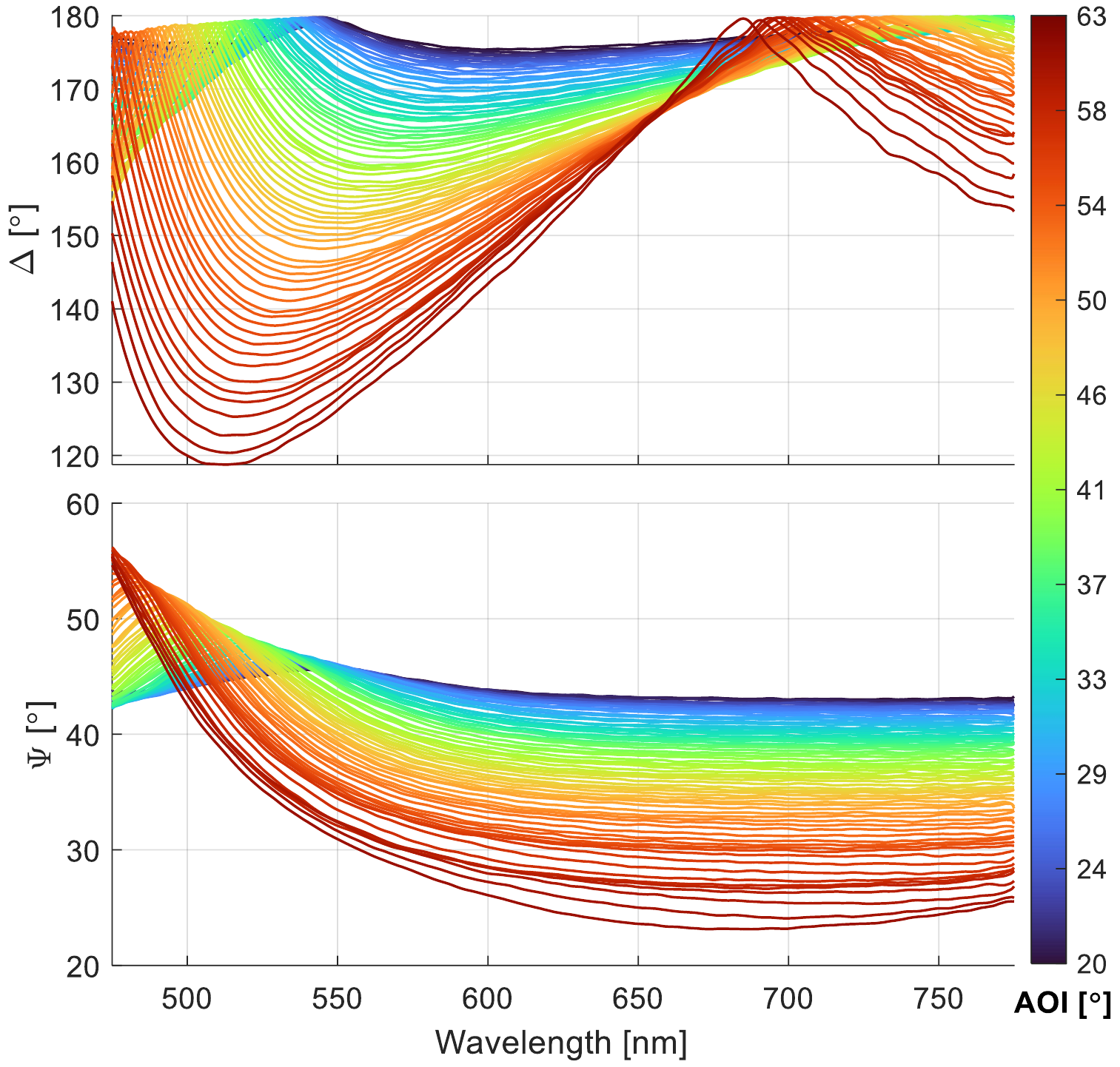}
     \caption{Spectrally and angularly resolved ellipsometry data ($\Psi$ and $\Delta$) obtained by the calibrated SME in a single-shot measurement on nominal 285 nm SiO$_2$ on Si; including 78 different AOIs between 20.0$^{\circ}$ and 62.5$^{\circ}$, and 601 wavelengths between 475 nm and 775 nm.}
\label{fig:285data}
\end{figure}

A single-shot data from the calibrated SME having a spot size of 5 microns is demonstrated here. The spectrally resolved $\Psi$ and $\Delta$ from a silicon wafer with nominal 285 nm oxide layer are measured at 78 different AOIs between 20.0$^{\circ}$ and 62.5$^{\circ}$ simultaneously, as seen in Figure \ref{fig:285data}. This is an example of ellipsometric data acquisition capability of the SME in a single-shot of few seconds, yielding spectrally and angularly resolved ellipsometric information from an area of a few microns. This, as far as known to us, makes the SME fastest in ellipsometric data acquisition.

The substantial increase in data acquisition rate gives the SME a practical advantage on sensitivity to fit parameters, increasing the accuracy of its fitting results. Currently, the SME can easily achieve fine resolution in both spectra and AOIs, with a spectral resolution less than 0.5 nm and an AOI resolution less than 0.5 degrees. The spectral and angular resolution can be further increased or varied as needed, depending on the system hardware.

\section{Raman spectra of MoS$_2$ and graphene flakes}
\label{ramansection}

The Raman spectra of the MoS$_2$ and graphene flakes which are measured by the SME in the paper's subsection \ref{2Dsection} are demonstrated here. The Raman spectra are measured using a 514.5 nm excitation laser and all measured flakes are on substrates with nominal 285 nm SiO$_2$ on Si.

\begin{figure}[H]
\centering
\includegraphics[width=1\textwidth]{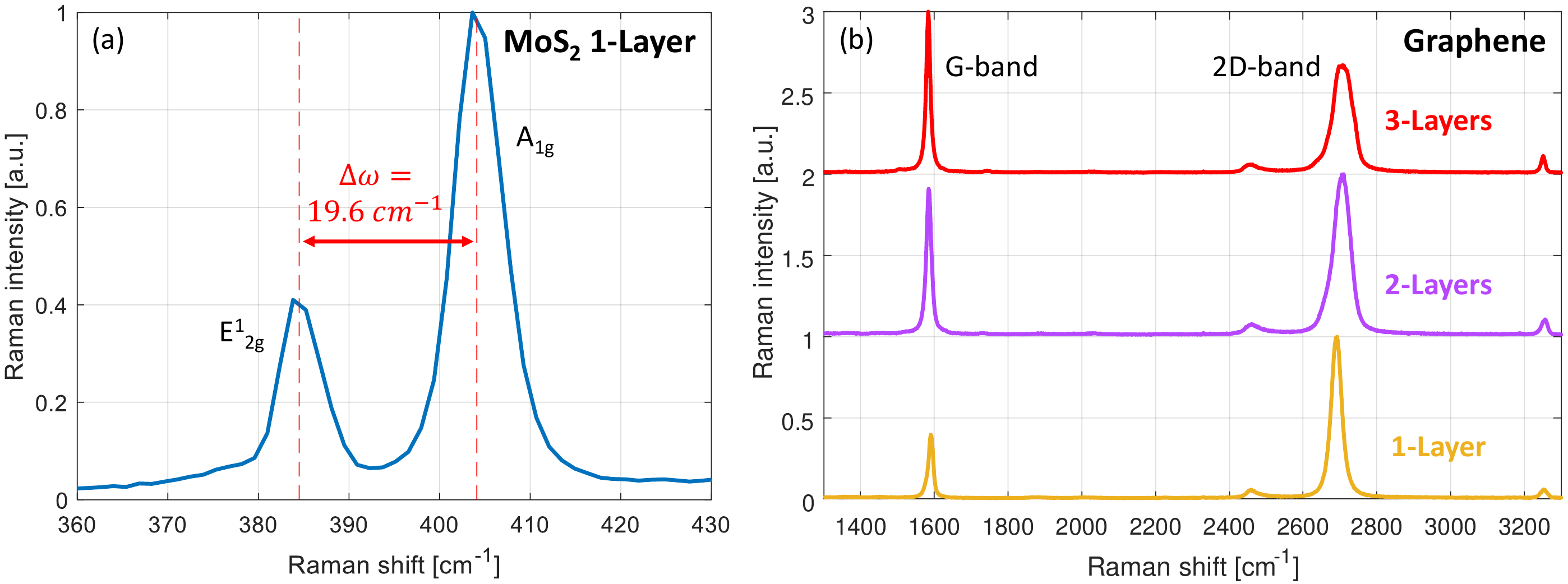}
     \caption{Raman spectra of (a) the monolayer MoS$_2$ flake and (b) the monolayer, bilayer and trilayer graphene flakes which are measured by the SME in subsection \ref{2Dsection} of the paper.}
\label{fig:ramans}
\end{figure}

Figure \ref{fig:ramans}a is the Raman spectra obtained on the MoS$_2$ flake measured by the SME for its complex refractive index. The in-plane E$^{1}_{2g}$ vibrational mode at 384.5 cm$^{-1}$ and out-of-plane A$_{1g}$ vibrational mode at 404.1 cm$^{-1}$ show a frequency difference of $\Delta \omega=19.6$ $cm^{-1}$ which is in excellent agreement with the literature on monolayer MoS$_2$ \cite{Lee2010AnomalousMoS2}, confirming the flake's monolayer nature.

Figure \ref{fig:ramans}b shows the Raman spectra of the monolayer, bilayer and trilayer graphene flakes which are measured by the SME for demonstration of its ellipsometric signal sensitivity to single-atom thick layers. The frequencies, shapes and peak ratios of the G and 2D band vibrational modes are in excellent agreement with the literature \cite{Wang2008RamanEffect}, confirming the mono-, bi- and trilayer nature of the graphene flakes.

\end{document}